\documentclass[showkeys,12pt, preprint,preprintnumbers,nofootinbib, groupedaddress,superscriptaddress,amsmath,amssymb]{revtex4}
\usepackage{graphicx}
\usepackage{dcolumn}
\usepackage{bm}
\usepackage{amssymb}
\usepackage{amsmath}
\usepackage{epsfig}    
\usepackage{color}
\usepackage{slashed}
\usepackage{hhline}

\def\be{\begin{equation}}
\def\ee{\end{equation}}
\newcommand{\bea}{\begin{eqnarray}}
\newcommand{\eea}{\end{eqnarray}}
\newcommand{\nn}{\nonumber}

\numberwithin{equation}{section}

\begin{document}

{\begin{flushright}{KIAS-P17013}
\end{flushright}}

\title{Radiatively Induced Neutrino Mass Model \\
with\\
Flavor Dependent Gauge
Symmetry  } 
\author{SangJong Lee }
\affiliation{School of Physics, KIAS, Seoul 02455, Korea}

\author{Takaaki Nomura}
\email{nomura@kias.re.kr}
\affiliation{School of Physics, KIAS, Seoul 02455, Korea}

\author{Hiroshi Okada}
\email{macokada3hiroshi@cts.nthu.edu.tw}
\affiliation{Physics Division, National Center for Theoretical Sciences, Hsinchu, Taiwan 300}

\date{\today}

\begin{abstract}
We study a radiative seesaw model at one-loop level with a flavor dependent gauge symmetry $U(1)_{\mu-\tau}$, in which we consider bosonic dark matter. 
We also analyze the constraints from lepton flavor violations, muon $g-2$, relic density of dark matter, and collider physics,
 and carry out numerical analysis to search for allowed parameter region which satisfy all the constraints and to investigate some predictions.   
Furthermore we find that a simple but adhoc hypothesis induces specific two zero texture with 
 inverse mass matrix, which provides us several predictions such as a specific pattern of Dirac CP phase.

\end{abstract}
\maketitle
\newpage

\section{Introduction}
The observation of neutrino oscillation confirms at least two non-zero masses of active neutrinos indicating physics beyond the standard model (SM) to generate the neutrino masses.
Radiative seesaw models are one of the attractive candidate to generate the neutrino masses where a neutrino mass matrix is induced at loop level and a dark matter (DM) candidate can be included as a particle propagating inside a loop diagram for generating neutrino mass. 
It is also interesting to include flavor dependent gauge symmetry with which we can obtain predictive structure of neutrino mass matrix~\cite{Crivellin:2015lwa,Kownacki:2016pmx}.

One of the interesting flavor dependent $U(1)$ gauge symmetry is the $U(1)_{\mu - \tau}$ which can induce sizable deviation of muon anomalous magnetic dipole moment from SM prediction, $\Delta a_\mu$, where experimental observation indicates $\Delta a_\mu \simeq O(10^{-9})$ suggesting discrepancy from the SM value~\cite{bennett}.
In addition, some interesting phenomenologies regarding the $U(1)_{\mu - \tau}$ are investigated, e.g. in Refs.~\cite{He:1990pn,Ma:2001md,Altmannshofer:2014cfa,Heeck:2014qea,Araki:2014ona,Araki:2015mya,Baek:2015fea,Baek:2015mna,Baek:2001kca,Chun:2007vh,Harigaya:2013twa,Baek:2008nz, Ko:2017quv, Ko:2017yrd, Altmannshofer:2016brv}.
{The $U(1)_{\mu - \tau}$ symmetry also can constrain the structure of  Majorana mass matrix of neutrinos giving predictability for neutrino sector.
However, it is not so trivial when the active neutrino mass matrix is generated via radiative seesaw mechanism.  }
We then apply the $U(1)_{\mu - \tau}$ gauge symmetry in a radiative seesaw model and investigate prediction in neutrino mass matrix.

In this paper, we construct a radiative seesaw model with $U(1)_{\mu - \tau}$ gauge symmetry and $Z_2$ symmetry in which we introduce exotic $SU(2)_L$ doublet leptons with $U(1)_{\mu -\tau}$,  a $Z_2$ even singlet scalar field, and $Z_2$ odd triplet and singlet scalar fields.  
In the model, active neutrino mass matrix is generated at one loop level where $Z_2$ odd particles propagate inside a loop diagram. 
Furthermore we have DM candidate which is the lightest $Z_2$ odd neutral particle.
Then global numerical analysis is carried out to search for allowed parameter region and to investigate some predictions in the model, taking into account constraints from charged lepton flavor violation (cLFV), $\Delta a_\mu$, and relic density of DM.
In addition, we find that structure of the Dirac mass matrix of exotic lepton determines that of the active neutrino mass matrix when we apply assumptions $i)$ degenerate masses for exotic leptons, or $ii)$ some vanishing Yukawa couplings which are associated with interactions among SM leptons, exotic leptons and exotic scalars.
In that case, we have two zero texture of the neutrino mass matrix which provides some predictions in neutrino oscillation experiments.

This paper is organized as follows. 
In Sec.~II, we introduce our model and discuss some phenomenologies such as neutrino mass matrix, lepton flavor violation, and some processes induced by $Z'$ interactions.
The numerical analysis is carried out in Sec.~III to search for parameter region satisfying experimental constraints and to obtain some prediction for neutrino mass matrix.
Finally we summarize the results in Sec.~IV.

 \section{Model, particle properties and phenomenology}
 \begin{widetext}
\begin{center} 
\begin{table}
\begin{footnotesize}
\begin{tabular}{|c||c|c|c|c|c|c|c|c|c|c|}\hline\hline  
 & \multicolumn{9}{c|}{Leptons} \\\hline
Fermions  ~&~ $L_{L_e}$ ~&~ $L_{L_\mu}$ ~&~ $L_{L_\tau}$ ~&~ $e_R$ ~&~ $\mu_R$ ~&~ $\tau_R$ ~&~ $L'_{e}$ ~&~ $L'_{\mu}$ ~&~ $L'_{\tau}$~
\\\hline 
$SU(3)_C$  & $\bm{1}$  & $\bm{1}$  & $\bm{1}$   & $\bm{1}$  & $\bm{1} $  & $\bm{1}$ & $\bm{1}$  & $\bm{1} $  & $\bm{1}$ \\\hline 
 $SU(2)_L$  & $\bm{2}$  & $\bm{2}$  & $\bm{2}$   & $\bm{1}$  & $\bm{1}$   & $\bm{1}$  & $\bm{2}$  & $\bm{2}$   & $\bm{2}$ \\\hline 
$U(1)_Y$  & $-\frac{1}{2}$ & $-\frac12$ & $-\frac12$  & $-1$ &  $-1$  &  $-1$  & $-\frac12$ &  $-\frac12$  &  $-\frac12$ \\\hline
 $U(1)_{\mu-\tau}$ & $0$  & $1$ & $-1$ & $0$  & $1$   & $-1$ & $0$  & $1$   & $-1$ \\\hline
$Z_2$  & $+$  & $+$ & $+$ & $+$ & $+$ & $+$& $-$ & $-$ & $-$ \\\hline
\end{tabular}
\caption{Field contents of fermions
and their charge assignments under $SU(2)_L\times U(1)_Y\times  U(1)_{\mu-\tau}\times Z_2$.}
\label{tab:1}
 \end{footnotesize}
\end{table}
\end{center}
\end{widetext}

\begin{table}[t]
\centering {\fontsize{10}{12}
\begin{tabular}{|c||c|c||c|c|c|}\hline\hline
&\multicolumn{2}{c||}{VEV$\neq 0$} & \multicolumn{2}{c|}{Inert } \\\hline
  Bosons  &~ $\Phi$  ~ &~ $\varphi$ ~ &~ $\Delta$   ~ &~ $S$ ~ \\\hline
$SU(2)_L$ & $\bm{2}$ & $\bm{1}$ & $\bm{3}$  & $\bm{1}$    \\\hline 
$U(1)_Y$ & $\frac12$ & $0$ & $1$  & $0$    \\\hline
 $U(1)_{\mu-\tau}$ & $0$  & $1$ & $0$ & $0$   \\\hline
$Z_2$ & $+$ & $+$  & $-$ & $-$ \\\hline
\end{tabular}%
} 
\caption{Field contents of bosons
and their charge assignments under $SU(2)_L\times U(1)_Y\times U(1)'\times Z_2$, where $SU(3)_C$ singlet for all bosons. }
\label{tab:2}
\end{table}

In this section, we introduce our model and discuss some phenomenologies.
As extra symmetries, local $U(1)_{\mu - \tau}$ and discrete $Z_2$ symmetries are added.
In the fermion sector, we introduce $SU(2)_L$ doublet vector like fermions $L'_{{e,\mu,\tau}}\equiv[N,E]^T_{e,\mu,\tau}$, and impose a flavor dependent gauge symmetry
 $U(1)_{\mu-\tau}$ as summarized in Table~\ref{tab:1}.
 Also $Z_2$ odd parity is imposed for this new fermion in order to discriminate the SM model leptons with $SU(2)_L$ and forbid the mixing between them.~\footnote{Notice here that the neutral component of $L'$ cannot be a DM candidate, because it is ruled out by the direct detection search via Z boson portal.}
In the scalar sector, we add an $SU(2)_L$ triplet inert scalar $\Delta$, real singlet inert scalar $S$, and singlet scalar $\varphi$ to the SM Higgs $\Phi$ as summarized in Table~\ref{tab:2}. Notice here the Higgs doublet $\Phi$ (that spontaneously breaks electroweak symmetry), the $SU(2)$ singlet field $\varphi$ (that spontaneously break $U(1)_{\mu-\tau}$ symmetry), have the vacuum expectation values (VEVs), which are respectively symbolized by $v/\sqrt2$, $v'/\sqrt2$, and $Z_2$ odd parity is also imposed for the inert scalars $\Delta$ and $S$ to forbid the tree level neutrino masses through VEVs. Therefore the lightest neutral scalar boson with $Z_2$ odd parity can be a DM candidate.

{\it Yukawa interactions and scalar potential}:
Under these fields and symmetries, the renormalizable Lagrangians for quark and lepton sector are given by 
\begin{align}
-{\cal L}_{L} = & \sum_{\ell=e,\mu,\tau} \left[y_{\ell} \bar L_{L_\ell}  \Phi \ell_R + y_{S_\ell}\bar L_{L_\ell} L'_{R_\ell}S +M_\ell \bar L'_{L_\ell} L'_{R_\ell} \right]
 \nn\\&
+y_{\Delta_1} \bar L_{L_e}^C (i\sigma_2)\Delta L'_{L_e}  + y_{\Delta_2} \bar L_{L_\tau}^C (i\sigma_2)\Delta L'_{L_\mu} + y_{\Delta_3} \bar L_{L_\mu}^C (i\sigma_2)\Delta L'_{L_\tau} 
\nn\\&
+ y_{E_1} \varphi^* \bar L'_{L_e} L'_{R_\mu} +  y_{E_2} \varphi \bar L'_{L_e} L'_{R_\tau} + {\rm c.c.},
\label{eq:lag-quark}
\end{align}
where  $\sigma_2$ is the second Pauli matrix, and again $L'\equiv [N,E]^T$.

We parametrize the scalar fields as 
\begin{align}
&\Phi =\left[
\begin{array}{c}
w^+\\
\frac{v+\phi+iz}{\sqrt2}
\end{array}\right],\ 
{\eta =\left[
\begin{array}{c}
\eta^+\\
\frac{\eta_R+i\eta_I}{\sqrt2}
\end{array}\right]} 
,\ 
\Delta =\left[
\begin{array}{cc}
\frac{\Delta^+}{\sqrt2} & \Delta^{++}\\
\Delta^{0} & -\frac{\Delta^+}{\sqrt2}
\end{array}\right],
\ \Delta_0=\frac{\Delta_R + i\Delta_I}{\sqrt2},\ \varphi=\frac{v'+\rho+i z'}{\sqrt2},
\label{component}
\end{align}
where $v~\simeq 246$ GeV is VEV of the Higgs doublet, and $w^\pm$, $z$, and $z'$ are respectively Nambu-Goldstone boson(NGB) 
which are absorbed by the longitudinal component of $W$, $Z$, and $Z'$ boson; $Z'$ boson comes from $U(1)_{\mu-\tau}$ gauge field.
Then we have two neutral  boson mass matrices $m^2_{\rho\phi}$ and $m^2_{S\Delta}$ in the basis of $[\rho,\phi]^T$ and $[S,\Delta_R]^T$, and these are diagonalized by $O_{a}^T m^2_{\rho\phi}O_{a}\equiv$Diag[$m_{h_1},m_{h_2}$] and $O_{\alpha}^T m^2_{S\Delta}O_{\alpha}\equiv$Diag[$m_{H_1},m_{H_2}$] respectively, where the mixing source of $O_\alpha$ arises from the nontrivial quartic coupling $\lambda_0 \Phi^T(i\sigma_2)\Delta^\dag \Phi S$ and each mass eigenstate can be written in terms of couplings of Higgs potential~\footnote{See Appendix in details.}.
Here we define mixing matrices as
\begin{align}
O_{a(\alpha)} =
\left[\begin{array}{cc} c_{a(\alpha)} & s_{a(\alpha)} \\ -s_{a(\alpha)} & c_{a(\alpha)} \end{array}\right],\quad
s_a=\frac{2\lambda_{\Phi\varphi} {v v'} }{m_{h_1}^2-m_{h_2}^2},\quad
s_\alpha=\frac{2\sqrt2\lambda_{0} {v^2} }{m_{H_1}^2-m_{H_2}^2},
\label{eq:mixing}
\end{align}
where $c(s)_{a(\alpha)}$ is the short-hand notation of $\cos(\sin)_{a(\alpha)}$.
Notice here that we assmue small mixing case $O_a\approx {\bf 1}$ in following analysis, which could however be an natural assumption because $s_a\lesssim {0.4} $ is indicated from the data of LHC experiment~\cite{hdecay, Chpoi:2013wga, Cheung:2015dta,Dupuis:2016fda}; therefore we take $m_\rho\approx m_{h_1}$ and $m_{\phi}\approx m_{h_2}\equiv m_{h_{SM}}$.

After the $\mu-\tau$ gauge symmetry breaking, {\it vector-like fermion mass matrix} can be written in the basis $[L'_{e},L'_{\mu},L'_{\tau}]^T$ as follows:
\begin{align}
M_{L'}\equiv \left[\begin{array}{ccc}  M_e & M_{e\mu} & M_{e\tau}  \\ M_{e\mu} & M_\mu & 0 \\ M_{e\tau} & 0 & M_\tau \end{array}\right],\label{eq:ML'}
\end{align}
where we {have simply assumed $M_{L'}$ to be a real symmetric matrix} and define $M_{e\mu}\equiv y_{E_1} v'/\sqrt2$ and $M_{e\tau}\equiv  y_{E_2} v'/\sqrt2$. Then $M_{L'}$ is diagonalized by orthogonal mixing matrix $V$ ($VV^T=1$) as 
\begin{align}
V^T M_{L'} V =D_N \equiv {\text{Diag.} }\left[M_1,M_2,M_3\right],\quad N_{{e,\mu,\tau}}=V N_{{1,2,3}},\label{eq:N-mix}
\end{align}
where $M_{1,2,3}$ is the mass eigenstate.

\begin{figure}[!hptb]
\begin{center}
\includegraphics[scale=0.4]{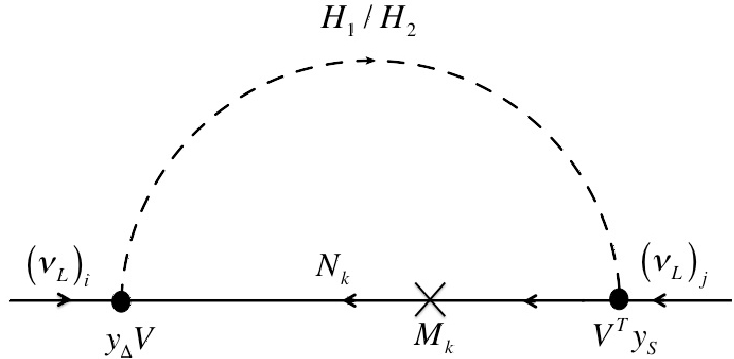}
\caption{The Feynman diagram for generating neutrino mass matrix.  } 
  \label{fig:neutrino}
\end{center}\end{figure}

\subsection{Active neutrino mass and lepton flavor violating processes}
{\it Our active neutrino mass matrix} is given in general at one-loop level by the diagram shown in Fig.~\ref{fig:neutrino} which is calculated as~\cite{Okada:2015vwh}
\begin{align}
m_\nu^{th}&= 
y_{\Delta} \epsilon V {D_N} R V^T y_S  + [y_{\Delta} \epsilon V {D_N} R V^T y_S]^T ,
\label{eq:neut-mass}\\
\epsilon &\equiv
\left[\begin{array}{ccc}  1 & 0 & 0 \\ 0 & 0 & 1 \\ 0 & 1 & 0 \end{array}\right],\quad 
{ R} =\frac{s_\alpha c_\alpha}{(4\pi)^2}
\left[\frac{r_{k_2}\ln r_{k_2}}{1-r_{k_2}}-\frac{r_{k_1}\ln r_{k_1}}{1-r_{k_1}}\right],
\end{align}
where $r_{k_i} \equiv(m_{H_i}/M_k)^2$, (i=1,2),  $y_S$ and $y_\Delta$ are diagonal Yukawa matrices respectively.
{Here we derived neutrino mass formula using mass eigenstates of scalar bosons and exotic fermions; then the vertices in the diagram are products of coupling $y_{\Delta, S}$ and mixing matrix $V$, and contribution from $\lambda_0$ coupling with the SM Higgs VEV insertions is included in the scalar mixing $s_\alpha$ as Eq.(\ref{eq:mixing}). }
On the other hand, the neutrino mass matrix can be written in terms of experimental values as  $m_\nu^{exp}=U D^\nu U^T$,
\footnote{In the current experiment, only five parameters are measured; two mass difference squared and three mixing angles.}
where $U$ is 3 by 3 unitary mixing matrix and $D^\nu\equiv {\rm diag.}[m_{\nu_1} e^{i\rho},m_{\nu_2} e^{i\sigma},m_{\nu_3}]$ is neutrino mass eigenvalues~\cite{Fritzsch:2011qv}. Therefore we have to satisfy the relation
$m_\nu^{th}\approx m_\nu^{exp}$.
{The smallness of neutrino masses $\sim10^{-12}$ GeV partly arises from loop suppression factor and small mixing of $s_\alpha\sim$0.1; $s_\alpha c_\alpha/(4\pi)^2\sim10^{-3}$, but the other factor is controlled by Yukawa couplings; $y_\Delta y_S\sim10^{-11}$ for $D_N\approx{\cal O}$(100) GeV. Thus each of Yukawa couplings could typically be the same order of electron Yukawa coupling; $y_\Delta \sim y_S\sim10^{-5}$. On the other hand the mass hierarchy between $M$ and $m_{H_a}$ does not severely affect the order of neutrino masses.}


{\it  Lepton flavor violations(LFVs)} arises from the term $y_\Delta$ and $y_S$ at one-loop level, and its form can be given by
\begin{align}
& {\rm BR}(\ell_i\to\ell_j\gamma)= \frac{48\pi^3\alpha_{\rm em} C_{ij} }{{\rm G_F^2} m_{\ell_i}^2}\left(|a_{R_{ij}}|^2+|a_{L_{ij}}|^2\right),\\
 a_{R_{ij}} &=\frac{1}{(4\pi)^2}
\sum_{k=1,2,3}\left[
\frac{Y^\dag_{\Delta_{ki}} Y_{\Delta_{jk}}}{2}\left(m_{\ell_j} F_2 [M_k,m_{\Delta^\pm}]
+2 m_{\ell_i} (2F_2[M_k,m_{\Delta^{\pm\pm}}]+F_2[m_{\Delta^{\pm\pm}},M_k]) \right)
\right.\nn\\
&- \left. Y^\dag_{S_{ki}} Y_{S_{jk}} m_{\ell_i} 
\left(c_\alpha^2 F_2 [H_1,M_k] + s_\alpha^2 F_2 [H_2,M_k]  \right)\right],\\
\quad 
a_{L_{ij}} &=\frac{1}{(4\pi)^2}
\sum_{k=1,2,3}\left[
\frac{Y^\dag_{\Delta_{ki}} Y_{\Delta_{jk}}}{2}\left(m_{\ell_i} F_2 [M_k,m_{\Delta^\pm}]
+2 m_{\ell_j} (2F_2[M_k,m_{\Delta^{\pm\pm}}]+F_2[m_{\Delta^{\pm\pm}},M_k]) \right)
\right.\nn\\
&- \left. Y^\dag_{S_{ki}} Y_{S_{jk}} m_{\ell_j} 
\left(c_\alpha^2 F_2 [H_1,M_k] + s_\alpha^2 F_2 [H_2,M_k]  \right)\right],\\
& F_{2}(m_a,m_b)=\frac{2 m_a^6 +3 m_a^4 m_b^2 -6 m_a^2 m_b^4 +6 m_b^6+12 m_a^4 m_b^2 \ln(m_b/m_a)}{12(m_a^2-m_b^2)^4},
\end{align}
where $Y_\Delta\equiv  y_\Delta\epsilon V$, $Y_S\equiv y_S V$, $\eta^\pm$ is the singly charged component of $\eta$, ${\rm G_F}\approx 1.17\times10^{-5}$[GeV]$^{-2}$ is the Fermi constant, $\alpha_{\rm em}\approx1/137$ is the fine structure constant, $C_{21}\approx1$, $C_{31}\approx 0.1784$, and $C_{32}\approx 0.1736$.
Experimental upper bounds are respectively given by ${\rm BR}(\mu\to e\gamma)\lesssim 4.2\times10^{-13}$, ${\rm BR}(\tau\to e\gamma)\lesssim 3.3\times10^{-8}$, and ${\rm BR}(\tau\to \mu\gamma)\lesssim 4.4\times10^{-8}$.
\\
{\it New contributions to the muon anomalous magnetic moment (muon $g-2$: $\Delta a_\mu$)} arises from Yukawa terms  $y_\Delta$ with negative contribution and $y_S$  with positive contribution.
Also another source via additional gauge sector can also be induced by
\begin{align}
&\Delta a_\mu =\Delta a_\mu^{Yukawa} +  \Delta a_{\mu}^{Z'},\\
&\Delta a_\mu^{Yukawa}=-m_\mu [a_R+a_L]_{\mu\mu},
\quad \Delta a_{\mu}^{Z'}\approx \frac{g_{Z'}^2}{8\pi^2}\int_0^1 da \frac{2 r a (1-a)^2}{r(1-a)^2+a}, \label{eq:G2-ZP}
\end{align}
where $r\equiv(m_\mu/M_{Z'})^2$, and $Z'$ is the new gauge vector boson. Thus we could explain the sizable muon $g-2$ $(\approx{\cal O}[10^{-9}])$~\cite{bennett}, if we can satisfy the constraint of trident process.
 Notice here that $g_{Z'}\lesssim10^{-3}$~\cite{Altmannshofer:2014pba} has to be satisfied due to the trident process.

\begin{figure}[!hptb]
\begin{center}
\includegraphics[scale=0.4]{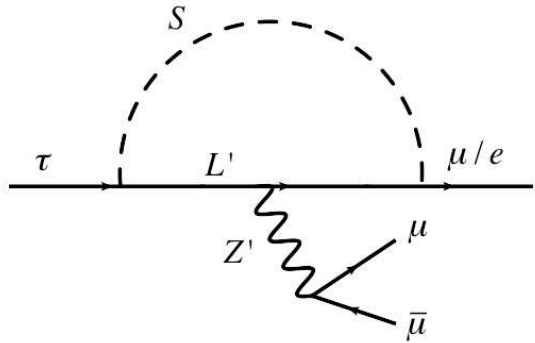}
\caption{The Feynman diagram for three body LFV decay via $Z'$.  } 
  \label{fig:LFV}
\end{center}\end{figure}
{
It is worthwhile to estimate three body decays; BR$(\tau\to\mu\bar\mu e)$ and BR$(\tau\to\mu\bar\mu\mu)$ via $Z'$ boson at one-loop level in Fig.~\ref{fig:LFV}; 
tree level contribution from $Z'$ is absent since $Z'$-charged lepton interactions are flavor diagonal.
Then our formula is evaluated by~\cite{Crivellin:2013hpa}
\begin{align}
{\rm BR}(\tau\to\mu\bar\mu \ell_\ell)\approx \frac{N_\ell m^5_\tau}{768\pi^3\Gamma_\tau}\left(\frac{g_{Z'}}{4\pi M_{Z'}}\right)^4
|G_2(y_S,V,M,m_S)|^2,
\end{align}
where $G_2\lesssim 0.1$ includes a loop function, $m_\tau\approx$1.777 GeV,   $\Gamma_\tau\approx2.3\times10^{-13}$ GeV, 
$N_\ell=1$ for $\ell=e$ and $N_\ell=1/2$ for $\ell=\mu$.
Once we put typical values into the above formula, one finds to be
\begin{align}
{\rm BR}(\tau\to\mu\bar\mu \ell)\approx{\cal O}(10^{-10}),
\end{align}
where we have adopted $g'/M_{Z'}\approx1/(400\ {\rm GeV})$ 
and $y_SV\approx{\cal O}(1)$. Since the experimental upper bounds are of the order $10^{-8}$~\cite{Olive:2016xmw}, our model does not restrict these modes for whole the parameters. 
}


\subsection{Dark matter}
Here we consider the lightest inert boson $X\equiv H_1\approx S$, assuming $s_\alpha<<1$ for simplicity.
{As we commented in previous subsection, this small mixing plays a role of suppression factor in neutrino mass formula while mass relation does not give significant change in the neutrino mass.}
Then annihilation modes generally arise from interactions associated with coupling constants $y_\Delta$ and $y_S$, and SM-Higgs portal. However
we found that Yukawa modes cannot explain the sizable relic density; 
Its cross section is of the order $10^{-10}$ GeV$^{-2}$ at most, even when large coupling $y_S$ is favor of the muon $g-2$. 
Thus we should rely on interactions in the scalar sector to explain thermal relic density of DM.
Then we focus on interactions between $h_1$ and $X$ since the SM Higgs portal interaction is highly constrained by the direct detection experiments.
Before considering the relic density, we also have to discuss the direct detection bound for $h_1$ portal coupling.
The stringent bound comes from spin independent nucleon-DM scattering via the ${h_1}$ scalar boson portal,
\footnote{Note here that the SM Higgs $h_2$ portal does not satisfy the relic density and direct detection simultaneously~\cite{Kanemura:2010sh} except the pole DM mass at $m_{h_2}/2$ and we assume $\lambda_{\Phi S}$ to be small to avoid constraints from direct detection.}
and its cross section is evaluated as
\begin{align}
\sigma_{SI}\approx (3.21\times10^{-29})\times \frac{m_N^4 {\lambda_{\varphi S}^2 s_a^2 c_a^2 }\ {\rm GeV^2}}{4\pi m_{h_1}^4(m_N+M_X)^2}\ {\rm cm^2},
\end{align}
where 
$\lambda_{\varphi S}$ is coefficient of $|\varphi|^2 S^2$, and $m_N\approx0.939$ GeV is neutron mass. The recent experiment LUX~\cite{Akerib:2016vxi} provides the bound on the scattering cross section as $\sigma_{SI}\lesssim 2.8\times10^{-46}$ cm$^2$ at $M_X\approx110$ GeV. This can be interpreted by the following bound
\begin{align}
 \lambda_{\varphi S} s_a \lesssim 0.021,
\end{align}
where we have used $ m_{h_1} = 125$ GeV as a reference value and $c_a \sim 1$ is assumed.
Hereafter we apply Max$[  s_a \lambda_{\varphi S}]= 0.021$ in the analysis of relic density.

{\it Relic density}:
The relevant annihilation cross sections to explain the relic density arise from the same coupling $\lambda_{\varphi S}$ in the scalar sector; $2X\to 2 {h_2}, 2Z,W^+W^-, t\bar t$. Note here that the other modes such as $2X\to b\bar b$ are sufficiently small than the dominant modes, since its related coupling of $b$ is of the order $10^{-2}$ at most.
Then the dimensionless cross section $W(s)$ is given by
\begin{align}
W(s) = 
& \frac{\lambda^2_{\varphi S}}{16\pi} \biggl(
\frac{ s_a^2  }{|s-m^2_{h_1} + im_{h_1} \Gamma_{h_1}|^2} 
\biggl[ \sum_{V=Z,W} 4m^4_V \sqrt{1-\frac{4m^2_V}{s}} \left(3-\frac{s}{m^2_V}  +\frac{s^2}{4m^2_V}\right )  \nonumber \\
& +6 m^2_t \sqrt{1-\frac{4m^2_t}{s}}(s-4m^2_t) \biggr] +
 \frac1\pi\sqrt{1-\frac{4m^2_{h_2}}{s}}\int d\Omega\left|  {s_a^2}  +\frac{\lambda_{\Phi\varphi}v_\varphi v}{4}
 \frac1{s-m^2_{h_1} + im_{h_1} \Gamma_{h_1}}\nn\right. \\
 &\left. 
 +3\frac{\lambda_{\Phi} {s_a} v^2}{2}
 \frac1{s-m^2_{h_2} + im_{h_2} \Gamma_{h_2}}
 +\frac{\lambda_{\varphi S} s_a^2 v^2}{4}\left(\frac{1}{t-M_X^2}+\frac{1}{u-M_X^2}\right) \right|^2
 \biggr),
\end{align}
where {$s,t,u$ are Mandelstam valuables, $c_a \simeq 1$ is taken}, we have assumed narrow width of $h_{1/2}$ as $\Gamma_{h_{1/2}}<<m_{h_{1/2}}$ GeV, and fixed $m_t\approx172.44$ GeV, $m_Z\approx91.2$ GeV, $m_W\approx80.4$ GeV.
Then the relic density of DM is given by~\cite{Edsjo:1997bg}
\begin{align}
&\Omega h^2
\approx 
\frac{1.07\times10^9}{\sqrt{g_*(x_f)}M_{Pl} J(x_f)[{\rm GeV}]},
\label{eq:relic-deff}
\end{align}
where $g^*(x_f\approx25) \approx 100$ is the degrees of freedom for relativistic particles at the freeze-out temperature $T_f = M_X/x_f$, $M_{Pl}\approx 1.22\times 10^{19}$ GeV,
and $J(x_f) (\equiv \int_{x_f}^\infty dx \frac{\langle \sigma v_{\rm rel}\rangle}{x^2})$ is given by~\cite{Nishiwaki:2015iqa}
\begin{align}
J(x_f)&=\int_{x_f}^\infty dx\left[ \frac{\int_{4M_X^2}^\infty ds\sqrt{s-4 M_X^2} W(s) K_1\left(\frac{\sqrt{s}}{M_X} x\right)}{16  M_X^5 x [K_2(x)]^2}\right]. \label{eq:relic-deff}
\end{align}
Then one has to satisfy the current relic density of DM; $\Omega h^2\approx 0.12$~\cite{Ade:2013zuv}.
 In our numerical analysis below we focus on annihilation mode of $2X \to \{ZZ, W^+W^-, t \bar t, 2h_2 \}$ assuming $m_{h_1}$ to be heavy.  
{Also we have assumed other scalar contact interactions such as $\lambda_{\Phi S}$ is small, and we have neglected mixing between $Z-Z'$, thus we do not consider the modes $2X\to Z'Z'$. }

\section{Numerical analysis}
\begin{figure}[!hptb]
\begin{center}
\includegraphics[width=70mm]{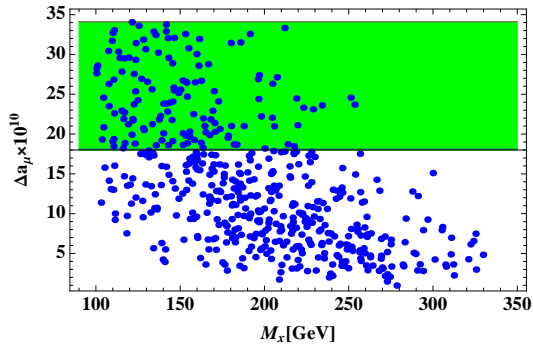} \qquad
\caption{Scattering plots in terms of muon $g-2$ and $M_X$ to satisfy the neutrino oscillation data and LFVs, where green region is in good agreement with the current experimental data $(26.1\pm8.0)\times10^{-10}$. } 
  \label{fig:DM-damu}
\end{center}\end{figure}

In this section, we show a global analysis,
where we have fixed some parameters for simplicity.
At first, we fix $m_{H_2}= m_{\Delta^\pm}= m_{\Delta^{\pm\pm}}$ in order to evade the constraints from oblique parameters in the triplet boson; the S, T, U-parameters are suppressed when the masses in the triplet are degenerated~\cite{Nomura:2016dnf}.
Also we numerically solve our parameters $Y\equiv (y_{S_e},y_{S_\mu},y_{S_\tau},y_{\Delta_2},y_{\Delta_3})$, by using the relation
$m_{\nu}^{th}=m_{\nu}^{exp}$,~\footnote{In principle, six parameters can numerically be solved, but it is technically difficult in our model.} where we impose the perturbative bounds on these output parameters; $Y\lesssim \sqrt{4\pi}$.
Thus we randomly select the following range of reduced input parameters as
\begin{align}
& 100\ {\rm GeV}\le M_X\ {\rm GeV},  \ m_{H_2} \in [1.2M_X,2500]\ {\rm GeV},
\nn\\&
 |y_{\Delta_1}|  \in[0.1, 4\pi],\  (\rho,\sigma)  \in[0, \pi],\  \delta \in[\pi, 2\pi],\  |s_\alpha| \in[10^{-5},0.1],\\
&  M_{Z'}\in [10^{-3},10^3]\ [{\rm GeV}],\  g_{Z'} \in[10^{-5},10^{-3}],
\end{align}
where we have used experimental neutrino oscillation data in ref.~\cite{Forero:2014bxa} with  3$\sigma$ range.
In Fig.~\ref{fig:DM-damu}, we show the scattering allowed plots in terms of muon $g-2$ and $M_X$ to satisfy the neutrino oscillation data and LFVs, where green region is in good agreement with the current experimental data $(26.1\pm8.0)\times10^{-10}$.
It shows that there is allowed region simultaneously to satisfy the muon $g-2$ and relic density of DM.

\begin{figure}[!hptb]
\begin{center}
\includegraphics[width=70mm]{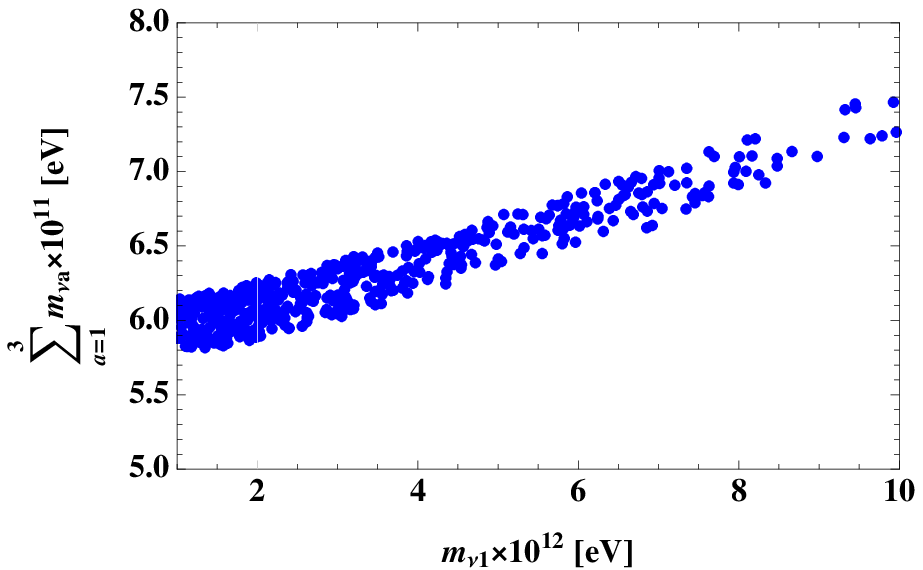} 
\caption{Scattering plots in terms of sum of neutrino masses and $m_{\nu_1}$ in the left panel and Dirac phase $\delta$ and Majorana phases; $\rho$(with red points) and $\sigma$(with blue points), in the right panel. } 
  \label{fig:neut-osci}
\end{center}\end{figure}

In Fig.~\ref{fig:neut-osci}, we show the allowed scattering plots in terms of sum of neutrino masses and $m_{\nu_1}$. It suggests that the lightest neutrino mass is of the order $10^{-12}$ eV.

\begin{figure}[!hptb]
\begin{center}
\includegraphics[width=70mm]{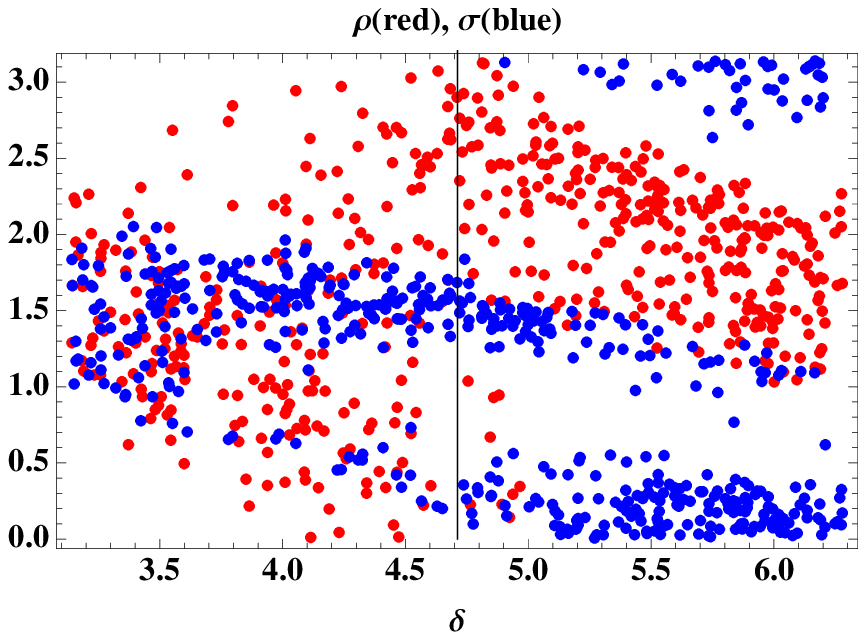}
\caption{Scattering plots in terms of sum of neutrino masses and $m_{\nu_1}$ in the left panel and Dirac phase $\delta$ and Majorana phases; $\rho$(with red points) and $\sigma$(with blue points), in the right panel. } 
  \label{fig:neut-phase}
\end{center}\end{figure}

In Fig.~\ref{fig:neut-phase}, we demonstrate Majorana phases; $\rho$(with red points) and $\sigma$(with blue points) in terms of Dirac phase $\delta$, where the red/blue present the region in $M_X\in[100,350]\ {\rm GeV}$.
\footnote{Here we take the upper bound 350 GeV on $M_X$. This is simply because the larger mass region than 250 GeV does not satisfy the sizable muon $g-2$ from loop diagram containing DM.}
It displays that $\delta$ runs over $\pi\sim 2\pi$, whereas Majorana phases tend to be localized, depending on $\rho$ and $\sigma$. Especially, both of these phases are in favor of being localized at around $\pi/2$ that could be  one of the remarkable features of this model.

\begin{figure}[t]
\begin{center}
\includegraphics[width=80mm]{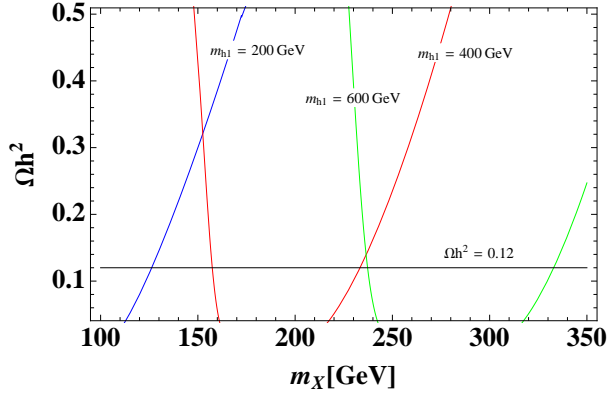} 
\caption{Plot line of relic density in term of the DM mass, where we have used  ${\rm Max} [\lambda_{\varphi S} s_a ]=0.02$, and 
horizontal line represents the measured relic density $\sim$0.12. Here the blue, red, and green line respectively represent the $h_1$ mass of 200 GeV, 400 GeV, and 600 GeV.} 
  \label{fig:relic-dm}
\end{center}\end{figure}

In fig.~\ref{fig:relic-dm}, we show the line of relic density in term of the DM mass, where we have used ${\rm Max} [\lambda_{\varphi S} s_a ]=0.02$, and horizontal line represents the measured relic density $\sim$0.12. Here the blue, red, and green line respectively represent the $h_1$ mass of 200 GeV, 400 GeV, and 600 GeV. Since the mass of $h_1$ is not constrained by any experiments discussed above, there are solutions in the whole mass range of DM that we have taken.


{\it Comment on the specific case}:
It is worth mentioning the following two hypotheses that lead a predictive two-zero texture with $(m_\nu)_{22}=(m_\nu)_{33}=0$:
 \begin{align}
&i)\ M\equiv  M_1\approx M_2\approx M_3,\\
&ii)\  y_{S_2}\approx y_{\Delta_3}\approx 0,\ {\rm or}\    y_{S_3}\approx y_{\Delta_2}\approx 0.
\end{align}
The case $i)$ suggests that a fermion DM is in a coannihilation system to satisfy the correct relic density of Universe when $M< m_{H_{1,2}}$. Notice here that the lower mass bound on $M$ is around 100 GeV from the LEP experiment. Transversely a bosonic DM candidate can simply satisfy the relic density.\\ 
The case $ii)$ suggests that a fermion DM does not require a coannihilation process among neutral fermions, however it must still be considered between the exotic charged fermions $E$ due to the constraint of oblique parameter.

\begin{figure}[!hptb]
\begin{center}
\includegraphics[width=75mm]{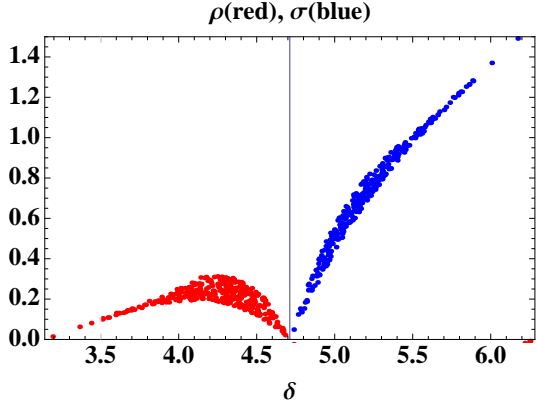} 
\caption{The allowed region between $\rho$(red) and $\sigma$(blue) in terms of $\delta$, where we adapt the recent global neutrino oscillation  data~\cite{Forero:2014bxa} up to 3$\sigma$ confidential level and the same input value in the general analysis.
It implies that the region of $\rho$ is restricted to be $0\sim3\pi/2$, whereas $\sigma$ $3\pi/2\sim 2\pi$,
and these are overlapped at around $\delta=3\pi/2$ that is the best fit value in the current neutrino experiments.
}   \label{fig:phases}
\end{center}\end{figure}

In both of the cases, the situation could be more or less same if we identify DM as the bosonic DM candidate,
and we adopt $i)$ in our discussion below.
Before starting  the discussion of neutrinos, let us roughly estimate the degree of our predictability from $\mu - \tau$ symmetry. 
Since we have eleven free parameters (three in $y_S$, three in $y_\Delta$, and five in $M_{L'}$) which contribute to form the texture, 
it still seems to remain nine free parameters even after imposing the above conditions $i)$ or $ii)$.
Thus naive expectation gives no predictions while one finds the type-C of neutrino texture that has only seven parameters.
It suggests that two more freedom in the parameter sets are reduced by our specific textures of $y_S$, $y_\Delta$, and $M_{L'}$ which are determined by the $\mu-\tau$ symmetry. Thus our model still improve predictability by two degrees of freedom due to the symmetry.~\footnote{Even if the general matrix case of $M_{L'}$ has eleven parameters (seven reals and four imaginaries), the two predictabilities does not changes, therefore one finds type-C of the neutrino texture due to reductions of eight parameters.}
%
Then $m_\nu$ is simplified as
\begin{align}
m_\nu &\approx R \left[y_{\Delta}\epsilon (V {D_N} V^T)y_{S} + y_{S}(V {D_N} V^T) \epsilon y_{\Delta}\right] \nonumber \\
& = R (y_{\Delta}\epsilon M_{L'}y_S+ y_SM_{L'}\epsilon y_{\Delta})\nn\\
&= R
\left[\begin{array}{ccc}  2 M_e y_{S_e}y_{\Delta_1} & M_{e\tau}y_{S_e}y_{\Delta_2}+M_{e\mu}y_{S_\mu}y_{\Delta_1} 
& M_{e\mu}y_{S_e}y_{\Delta_3}+M_{e\tau}y_{S_\tau}y_{\Delta_1}  \\  M_{e\tau}y_{S_e}y_{\Delta_2}+M_{e\mu}y_{S_\mu}y_{\Delta_1}  & 0 & M_\mu y_{S_\mu}y_{\Delta_3}+M_\tau y_{S_\tau}y_{\Delta_2} \\ 
M_{e\mu}y_{S_e}y_{\Delta_3}+M_{e\tau}y_{S_\tau}y_{\Delta_1} &  M_\mu y_{S_\mu}y_{\Delta_3}+M_\tau y_{S_\tau}y_{\Delta_2} & 0 \end{array}\right]\label{eq:type-c}\\
&\approx
\left[\begin{array}{ccc}  0.022-0.75 & 0.028-0.039 & 0.030-0.040  \\
0.028-0.039 & 0 & 0.023-0.75 \\ 
0.030-0.040 &  0.023-0.75  & 0 \end{array}\right] [{\rm eV}]
, \label{eq:exp-neut-oscil}
\end{align}
where we have used $V {D_N} V^{T} =  V V^T M_{L'} V V^{T}= M_{L'}$.
Eq.~(\ref{eq:type-c}) corresponds to the type C two zero texture that provides several predictions that only an inverted neutrino mass ordering is allowed and specific pattern of phases.
 In fig.~\ref{fig:phases}, we show $\rho$(red) and $\sigma$(blue) in terms of $\delta$, where we adapt the recent global neutrino oscillation  data~\cite{Forero:2014bxa} up to 3$\sigma$ confidence level and the same input value in the general analysis.
It implies that the region of $\rho$ is restricted to be $0\sim3\pi/2$, whereas $\sigma$ be $3\pi/2\sim 2\pi$,
and these are overlapped at around $\delta=3\pi/2$ that is in good agreement with the current neutrino experiments as the best fit value. In this case, the dominant contribution of muon $g-2$ arises from $\Delta a^{Z'}_\mu$, where $g_{Z'}\lesssim10^{-3}$~\cite{Altmannshofer:2014pba} is satisfied due to the trident process. While the relic density of DM can be obtained by the
Yukawa coupling $y_{S}$ that leads to the d-wave dominant.
This result is opposite to the one of general feature, although we do not show the detailed analysis here because this is nothing but ad-hoc hypothesis.

\section{ Conclusions and discussions}
We have proposed a radiative seesaw model at one-loop level with a flavor dependent gauge symmetry $U(1)_{\mu-\tau}$, in which we have consider gauge singlet-like bosonic dark matter candidate and explained muon $g-2$ without conflict of LFVs.
In the numerical analysis, we have shown several features as follows:
\begin{enumerate}
\item Whole the DM mass region with $\lambda_{\varphi S} s_a \approx 0.021$ is obtained by the experimental bounds on spin independent scattering and relic density of DM. And this range is in good agreement with the current experimental data of muon $g-2$ without conflict of LFVs as well as neutrino oscillation data. 
\item The typical lightest neutrino mass is of the order $10^{-12}$ eV.
\item There exist a mild correlation between the Dirac phase $\delta$ and Majorana phases $\rho,\sigma$.
Therefore, $\delta$ runs over $\pi\sim 2\pi$, whereas Majorana phases tend to be localized, depending on $\rho$ and $\sigma$. Especially, both of these phases are in favor of being localized at around $\pi/2$.

\item As a specific case such as $M\equiv  M_1\approx M_2\approx M_3$, we have found the predictive two zero texture(type-C)
and their features are clearer than the generic one. As an example, the region of $\rho$ is restricted to be $0\sim3\pi/2$, whereas $\sigma$ be $3\pi/2\sim 2\pi$,
and these are overlapped at around $\delta=3\pi/2$ that is in good agreement with the current neutrino experiments as the best fit value.
\end{enumerate}

Finally, we have an inert doubly charged Higgs boson which decay into dark matter and SM fermions by cascade decay modes.
It will be interesting to search for the signal of "missing $E_T$ + same sign leptons" as a signature of the inert Higgs triplet as well as our dark matter. 
The detailed analysis of the signal is beyond the scope of this paper and it will be studied elsewhere.  

{
If global $U(1)_{\mu-\tau}$ symmetry is applied to our model, a few results could change.
The first one is that the muon $g-2$ due to the absence of $Z'$ contribution.
The second one is that a new annihilation mode  of DM relic density has to be added; $2X\to 2G$, where $G$ is a physical massless goldstone boson. As a result, the allowed range of DM mass is wider, since whole the cross section increases.
}

\section*{ Appendix}
Here we give the most general  Higgs potential in a renormalizable theory as
\begin{align}
{\cal V}&=m_\Phi^2 \Phi^\dag\Phi+m_\varphi^2 \varphi^*\varphi+m_\Delta^2 {\rm Tr}[\Delta^\dag\Delta]+\frac12 m_S^2S^2\nn\\
&+(\lambda_0 \Phi^T(i\sigma_2)\Delta^\dag \Phi S+{\rm c.c.})
+\lambda_\Phi |\Phi^\dag\Phi|^2 + \lambda_\varphi |\varphi^*\varphi|^2  + \lambda_\Delta ({\rm Tr}[\Delta^\dag\Delta])^2  
+ \lambda_\Delta' {\rm Det}[\Delta^\dag\Delta] + \frac1{4!}\lambda_SS^4\nn\\
&+\lambda_{\Phi\varphi}(\Phi^\dag\Phi)( \varphi^*\varphi)
+\lambda_{\Phi\Delta}(\Phi^\dag\Phi) {\rm Tr}[\Delta^\dag\Delta]
+\lambda_{\Phi\Delta}'\sum_{i=1-3}(\Phi^\dag\tau_i\Phi) {\rm Tr}[\Delta^\dag\tau_i\Delta]
+\frac12\lambda_{\Phi S}(\Phi^\dag\Phi)S^2\nn\\
&+\lambda_{\varphi \Delta}\varphi^*\varphi  {\rm Tr}[\Delta^\dag\Delta]
+\frac12\lambda_{\varphi S}\varphi^*\varphi S^2
+\frac12\lambda_{ \Delta S} {\rm Tr}[\Delta^\dag\Delta] S^2.
\end{align}


\section*{Acknowledgments}
\vspace{0.5cm}
H. O. is sincerely grateful for all the KIAS members.

\end{document}